\documentclass[aps,amsmath,amssymb,preprintnumbers,groupedaddress,twocolumn]{revtex4}
\usepackage{graphicx}
\newcommand{\nc}{\newcommand}
\nc{\beq}{\begin{equation}} \nc{\eeq}{\end{equation}} \nc{\bea}{\begin{eqnarray}}
\nc{\eea}{\end{eqnarray}}

\def\gsim{\mathrel{\rlap{\lower4pt\hbox{\hskip1pt$\sim$}}
    \raise1pt\hbox{$>$}}}       %greater than or approx. symbol

\def\K3{{\bf K3}}

\def\n2d{\cN_{V^*}^{\otimes 2}}

\def\cN{{\mathcal N}}

\begin{document}

\preprint{UPR-1193-T}

\title{D-Instanton Generated Dirac Neutrino Masses}

\author{Mirjam Cveti{\v c$^{1         }$}}
\email{cvetic@cvetic.hep.upenn.edu}
\author{Paul Langacker$^{2         }$} \email{pgl@ias.edu}

\affiliation{$^{1         }$Department of Physics and Astronomy, University of Pennsylvania,
Philadelphia, USA \\$^{2         }$Institute for Advanced Study, Princeton, New Jersey, U.S.A.}

\begin{abstract}
\noindent{We present a stringy mechanism to generate Dirac
neutrino  masses by D-instantons in an experimentally relevant mass scale without fine-tuning. Within Type
IIA  string theory with  intersecting D6-branes, we spell
out specific conditions for the  emergence of such couplings and provide
a class of  supersymmetric local $SU(5)$ Grand Unified  models,
based on the $Z_2\times Z_2'$ orientifold compactification,  where
perturbatively absent Dirac neutrino masses  can be generated by
D2-brane instantons in the
experimentally observed mass regime, while Majorana masses remain absent,
thus providing  an intriguing  mechanism for the
origin of small  neutrino masses  due to non-perturbative stringy effects.}

\end{abstract}

%\pacs{}

\maketitle

%\date{today}

\bigskip

\section{Introduction}
Until recently, no satisfactory mechanism was known for generating
either Majorana neutrino masses (for a see-saw mechanism)
or small Dirac masses within the  Type IIA string theory context.
Recent work
\cite{Blumenhagen:2006xt,Ibanez:2006da} has shown that
Majorana masses can be generated non-perturbatively by D-brane instantons.
In this letter we show
that, as an equally plausible alternative, D-brane instantons may
instead generate small Dirac neutrino
masses at the observed scale without fine-tuning.

In the past year there have been  intriguing insights  into D-brane instantons
\cite{Blumenhagen:2006xt,Ibanez:2006da,Haack:2006cy,Florea:2006si}  which
can generate pertubatively absent
couplings  of genuinely string theoretic origin with apparently no field theory analogs. (For reviews
on the subject, see \cite{Akerblom:2007nh},\cite{Cvetic:2007sj}.) In type II string compactifications
with D-branes the specific superpotential couplings are typically forbidden due to perturbatively
conserved ``anomalous'' $U(1)$ factors. However, under specific conditions, that ensure the correct
number of D-instanton fermionic zero modes, these couplings can be generated with a  strength that is
exponentially suppressed by the D-instanton action.  The mechanism was originally proposed for
generating Majorana neutrino masses, the $\mu$-parameter and $R-parity$ violating terms
\cite{Blumenhagen:2006xt,Ibanez:2006da},  as well as the study of supersymmetry breaking effects
\cite{Florea:2006si}. Further efforts in these directions focused on  rational conformal field theory
searches  for global models with non-perturbatively realised Majorana masses \cite{Ibanez:2007rs}, an
explicit calculation of non-perturbative  Majorana neutrino couplings  within local orbifold
constructions \cite{Cvetic:2007ku} and further studies of  phenomenological implications for neutrino
physics \cite{Antusch:2007jd},  as well as  spelling out conditions for non-perturbatively induced
Yukawa couplings $10\,10\,5_H$ within SU(5)  Grand Unified Models (GUT's) \cite{Blumenhagen:2007zk}.
There have also been further analyses of  non-perturbative  supersymmetry breaking  effects
\cite{Aharony:2007db,Aharony:2007pr}.
% Within Type IIB framework,
%first global semi-realistic examples with non-perturbatively generated  Majorana masses and
%Polonyi-type supersymmetry breaking progress, have been constructed in \cite{CW}.

While local realisations of  models where this  D-instanton
 mechanism
% that  can realize
%certain non-perturbative couplings, such as Majorana neutrino couplings,
 were found
\cite{Cvetic:2007ku}, an important challenge was
in the construction of globally
 consistent semi-realistic  string vacua which  realize such
non-perturbative effects (for efforts within semi-realistic Gepner models, see
\cite{Ibanez:2007rs}). The main difficulty seems to have been due to
the specific Type IIA frameworks,
where conformal field theory techniques are applicable.
%There,  a tension between global
%consistency conditions in the D-brane sector and specific
%constraints on the allowed fermionic zero
%modes in the instanton background typically prevented the construction of globally
%consistent models.
On the other hand, the T-dual Type I framework with
magnetized
D-branes, described by stable holomorphic bundles on
compact Calabi Yau spaces, is amenable to
algebraic geometry techniques.
There, the  first classes  of semi-realistic globally consistent string
vacua,  where hierarchical couplings are  realised by D1-brane instantons,
were constructed
\cite{Cvetic:2007qj}.
% The  specific realizations employ elliptically
%fibered Calabi-Yau spaces with the role of D-instantons played by
%Euclidean D1-branes, wrapping holomorphic curves on Calabi-Yau
%space.

%Until recently, within the D-brane string
%constructions the origin of small neutrino masses was not well understood.
The purpose of this paper is to  present a new  mechanism
for small neutrino masses.
%, thus shedding light on the important problem of the origin of small neutrino masses in string theory.
Specifically,
we present D-brane vacua with a Standard Model
sector where perturbatively absent  Dirac neutrino masses
 are  generated non-perturbatively by D-instantons at a
desired mass scale without fine-tuning, while at the same time
ensuring the absence of non-perturbatively generated
Majorana neutrino masses. This string mechanism
 %can thus shed light on the  origin of neutrino and it
 should be contrasted
 with that for Majorana masses  \cite{Blumenhagen:2006xt,Ibanez:2006da}.
Both types of masses can be  generated by D-instantons that satisfy
specific conditions and are thus restricted to specific
 string vacuum solutions. However,
unlike D-instanton generated  Dirac neutrino masses,
the  desired mass scale for Majorana  neutrino masses
 can be achieved only after some fine-tuning of the  volumes of the cycles wrapped by D-instantons.

 For the sake of concreteness we
focus on the Type IIA framework with intersecting D6-branes, though the T-dual  formulation of
conditions in the Type I framework with magnetized D9-branes on Calabi-Yau spaces can also be described employing techniques developed in \cite{Cvetic:2007qj}. In these cases the D-instantons
 can generate exponentially suppressed Dirac neutrino masses at experimentally relevant mass scales, while the Majorana masses are not generated. Thus the proposal provides  a non-perturbative string realisation  of the origin of small Dirac neutrino  masses in the absence of a see-saw
 mechanism~\footnote{Small
 Dirac masses in supersymmetric field theory may also be generated by higher dimensional operators in the superpotential~\cite{Cleaver:1997nj,Langacker:1998ut} or K\"ahler potential~\cite{Demir:2007dt}, though this has not yet been implemented in a string construction. For a review, see~\cite{Langacker:2008yv}. If such terms were present in the proposed D-brane constructions they would merely be additive. Such higher dimensional operators  have  not been studied systematically in these classes of constructions.}.

 As a  concrete application we  construct a class of  local models, based on the $Z_2\times Z_2'$ toroidal orientifold,  which explicitly
realise this scenario.
Within this local construction we do not address the moduli stabilization issue,  a difficult problem in string theory. The  
back-reaction of supergravity fluxes and/or strong gauge dynamics, responsible to for moduli stabilization, can also affect  quantitative
results  for the proposed non-perturbative couplings (as well as global consistency constraints); however, this is beyond the scope of this paper.

The proposal is attractive since the classical instanton action enters the coupling at the
exponentially suppressed level, proportional to the inverse string coupling and the volume of the
cycles wrapped by the D-brane instanton.  These couplings are thus generically
 extremely small, and thus generate tiny Dirac neutrino masses
  without any additional tuning of
  the volume of the cycles.
  % wrapped by the D-instantons.
  This  mechanism should be contrasted
 with the case of D-instanton generated Majorana neutrino
  masses \cite{Blumenhagen:2006xt,Ibanez:2006da}  and  some other non-perturbatively
generated couplings, such as the $10\, 10\, 5_H$ Yukawa coupling of the SU(5) GUT  \cite{Blumenhagen:2007zk}, where some fine-tuning of the volumes of the cycles wrapped by D-instantons is needed in order to obtain the couplings in the desired regime, e.g. for Majorana masses in $10^{12-14}$ GeV range.

%constrained by the D-brane sector supersymmetry conditions. Thus, even before tackling full moduli
%stabilisation,  the desired range of exponentially suppressed couplings has to be compatible with the
%supersymmetry constraints.Our first class of examples realizes intermediate scale Majorana
%masses. In the second class of examples instantons generate a Polonyi-type superpotential in a hidden
%sector which breaks supersymmetry at an exponentially suppressed scale. These techniques can readily
%be applied to realize other instanton effects such as the generation of $\mu$-terms
%\cite{Blumenhagen:2006xt,Ibanz:2006da} or certain GUT Yukawa couplings \cite{Blumenhagen:2007zk}.

The specific focus are ``semi-realistic'' constructions within  the Type IIA string theory framework with
intersecting D6-branes wrapping three-cycles in the internal space (for  reviews, see
\cite{Blumenhagen:2005mu,Blumenhagen:2006ci}.).  Concrete realizations
will be based on Grand
Unified SU(5) models
\cite{Blumenhagen:2001te,Cvetic:2001nr,Cvetic:2002pj}, with chiral  families of quarks and leptons. However, again, we do not address the moduli stabilization. 
Specifically, the string
vacuum constructions should have the property that the Yukawa couplings responsible for Dirac neutrino
masses are absent perturbatively
%.  Namely, such  couplings are forbidden perturbatively
due to global
$U(1)$ selection rules. Focusing on Type IIA theory, under suitable circumstances Euclidean
D2-branes (E2-instantons) wrapping three cycles in the internal space  can break these global $U(1)$
symmetries to certain discrete subgroups and generate $U(1)$ violating interactions, as spelled out
in  \cite{Blumenhagen:2006xt,Ibanez:2006da,Florea:2006si}.

The Type IIA framework  allows for a geometric formulation of the zero mode conditions in the
presence of E2-instantons wrapping  rigid  three-cycles, and thus  an explicit geometric
interpretation of these non-perturbative string effects.  Namely, the intersection numbers of the of instanton  and D-brane cycles, which specify the number of charged fermionic zero modes, are topological numbers. However, in order to
illustrate the effects explicitly,  we write   expressions for these intersection numbers for a concrete local construction, based  the $Z_2\times Z_2'$  toroidal orientifold, with Hodge numbers $(h_{11}, h_{12}) = (3,51)$. We  employ the notation of
\cite{Blumenhagen:2005tn} (see also \cite{Dudas:2005jx}), to which we refer for details of the geometry and the construction of
rigid cycles. The orbifold group is generated by $\theta$ and $\theta'$ acting as reflections in the
first and last two tori, respectively. The $O6$-plane parallel to the instanton is an $O6^+$ plane,
whereas the three others are $O6^-$ planes.

The proposed framework requires  three stacks $a$, $b$ and $c$ of D6-branes giving rise to a
$U(5)_a\times U(1)_b\times U(1)_c$ gauge symmetry.
 The $U(5)_a$ splits into $SU(5)_a\times U(1)_a$, where the anomalous
$U(1)_a$ gauge boson gets massive via the generalized Green-Schwarz mechanism and
$U(1)_a$ appears as a global symmetry in
the effective action. The matter transforming as $\bf{10}$ under $SU(5)_a$ arises at intersections of
stack $a$ with its image $a'$; the matter fields transforming as $\bf{\bar{5}}$ as well as Higgs
fields $\bf{5}_H$ and $\bf{\bar{5}}_H$ are located at intersections of stack $a$ with $b$ and $b'$ or
$c$ and $c'$.  The Abelian gauge group factors  $U(1)_b$  and $U(1)_c$ also  acquires  massive gauge
bosons due to the generalized Green-Schwarz mechanism.

The key input in  the construction of the local model is summarized in Tables \ref{wn} and
\ref{sp}. Table \ref{wn}   lists the wrapping numbers  of the   D6-branes wrapping bulk
three-cycles $\Pi^B_{a}$, the building blocks of the local GUT models, and  the wrapping numbers of
the rigid three-cycle $\Pi_{\Xi}$ of the E2-instanton with the required zero mode structure.
%For a globally consistent model the concrete wrapping numbers decide if a combination of both abelian
%groups remains massless. If not, the model is of the usual Georgi-Glashow type, while in the presence
%of a massless $U(1)_X$ it represents a flipped $SU(5)$ model, whose intersection numbers are
%displayed in table \ref{tablegut}.
\begin{table}[ht]
\centering
\begin{tabular}{|c|c|c|c|}
\hline
 nos &  $(n_a^1,m_a^1)$& $(n_a^2,m_a^2)$&$(n_a^3,m_a^3)$   \\
\hline \hline
$5_a$  & $(\nu_2, 2\nu_2/n_2)$   & $(1,1)$ &  $(0,-1)$  \\
$1_b$  & $(n_2,2)$   & $(-1,2)$ &  $(-1,1)$  \\
$1_c$  & $(-1,0)$ & $(1,1)$ &  $(-1,1)$    \\
$1_E$  & $(1,0)$ & $(0,1)$ &  $(0,-1)$    \\
\hline
\end{tabular}
\caption{Wrapping numbers of D6-branes wrapping bulk three-cycles and
the wrapping numbers of the E2-instanton wrapping a rigid, orientifold
action invariant, three-cycle  on the $Z_2\times Z_2'$ toroidal
orientifold.  The wrapping numbers   $n_2$  and $\nu_2$ are positive integers, co-prime with 2 and
$2\nu_2/n_2$ integer, respectively.  The simplest choice is $n_2=\nu_2=1$. Another interesting
(non-Abelian anomaly free) case is   $n_2=\nu_2=3$.   The model is supersymmetric for the choice
$\chi_1=\chi_2=1/\sqrt{5}$, $\chi_3= n_2 \sqrt{5}/2$
of the complex structure toroidal moduli $\chi_i\equiv (R_2/R_1)_i$ of the $i^{th}$ two-torus.
\label{wn} } % \vspace{3mm}
\end{table}

\begin{table}[ht]
\centering
\begin{tabular}{|c|c|c|}
\hline
sector & number &  $U(5)_a\times U(1)_b\times U(1)_c$   \\
\hline \hline
$(a,a')$ &  $4(\nu_2-2\textstyle{\frac{\nu_2}{n_2}})$ & ${\bf {{ 10}}}_{(2,0,0)} + {\overline{\bf 15}}_{(-2,0,0)} $     \\
$(a,b')$ &  $16\nu_2$ & ${\bf 5}_{(1,1,0)}$    \\
$(a,c')$ &  $16 \textstyle{\frac{\nu_2}{n_2}}$ & ${\bf \overline{5}}_{(-1,0,-1)}$      \\
$(b,c')$ &  $16$ & ${\bf 1}_{(0,1,1)}$  \\
\hline
\end{tabular}
\caption{Chiral matter spectrum for the local $SU(5)$ GUT models  with intersecting D6-branes  on
the $Z_2\times Z_2'$ orientifold. The wrapping numbers of the D6-branes are depicted in  Table \ref{wn}. A
special case $\nu_2=n_2=3$ corresponds to the four family example and is free of non-Abelian
anomalies. Another four family example corresponds to $\nu_2=n_2=1$; however, in this case additional
``filler'' branes would have to be added to cancel  non-Abelian anomalies.}
\label{sp}  % \vspace{3mm}
\end{table}

We build a local  model on the $Z_2\times Z_2'$ orientifold  by wrapping $D6$ branes on the bulk
cycles specified by wrapping numbers $(n_i^a,m_i^a)$  (for further details see
\cite{Blumenhagen:2005tn}). 

The intersection numbers in  the respective $(a,b)$ and $(a,b')$ sectors are:
\begin{equation} I_{ab}=4 \prod_i(n_i^am_i^b-n_i^bm_i^a), \  I_{ab'}=-4
\prod_i(n_i^am_i^b+n_i^bm_i^a)\end{equation}
where  we  choose a convention that the chiral
superfields in the $a,{\bar b}$  representation correspond to $I_{ab}<0$. The symmetric and
anti-symmetric representations arise from the $(a,a')$ sector:
\begin{equation}I^{antisymm}=\textstyle{\frac{1}{2}}(I_{aa'}+I_{aO6}),\ \ \
I^{symm}=\textstyle{\frac{1}{2}}(I_{aa'}-I_{aO6}),
\end{equation}
 where  \begin{eqnarray}
 I_{aa'}&=&-32n_1^am_1^an_2^am_2^an_3^am_3^a\\
 I_{aO6} &=& -8m_1^am_2^am_3^a+8n_1^an_2^am_3^a-8m_1^an_2^an_3^a+8n_1^am_2^an_3^a \nonumber
 \end{eqnarray}
 As shown in \cite{Cvetic:2007ku}, the background of this model
exhibits one class of so-called rigid $O(1)$ instantons.
%is also shown in
% Table \ref{wrappingnumber}.
For completeness, in the conventions of \cite{Blumenhagen:2005tn},
we give the full
expression for the rigid three-cycle wrapped by the E2-instanton \bea
\Pi_{\Xi}&=&\frac{1}{4}\Pi^B_{\Xi}-\frac{1}{4}\sum_{i,j\epsilon(13)\times(12)}
\alpha^{\theta}_{ij,m} \\
&&+ \frac{1}{4}\sum_{j,k\epsilon(12)\times(12)} \alpha^{\theta'}_{jk,n}+
\frac{1}{4}\sum_{i,k\epsilon(13)\times(12)} \alpha^{\theta\theta'}_{ik,m}, \nonumber \label{E2} \eea
where  $\Pi^B_{\Xi}$ wraps the cycle $[(1,0)\,(0,1)\,(0,-1)]$. The twisted three-cycles  $\alpha^{\theta}_{jk,n}$ [$\alpha^{\theta}_{jk,m}$] can be understood as a product one-cycle $[a]^{I^\theta}$ ($[b]^{I^\theta}$) on the $I^\theta$-th two-torus and a two-cycle  $S^2$- a blow-up of  $(i,j) \in (1,2,3,4)\times (1,2,3,4)$  orbifold fixed points. (For further details see \cite{Blumenhagen:2005tn}.) The intersection number in the $(a,E)$
specifies the number of chiral fermionic modes and is of the form: \begin{equation}
 I_{aE}=
\prod_i(n_i^am_i^E-n_i^Em_i^a)\, .\end{equation} Again the convention $I_{aE}<0$ corresponds to
chiral fermionic zero modes in the $(a, -1_E)$ representation. Note that since D6-branes wrap (non-rigid) bulk three-cycles, the intersection number in the $(a,E)$
sector does not depend on fractional twisted three-cycles of the instanton.

The supersymmetry conditions are ensured by requiring that the three-cycles are special Lagrangians
with respect to the same holomorphic three-form. In the case of toroidal compactification these take
the form:
\begin{eqnarray}
m_1^am_2^am_3^a&-&\sum_{i\ne j\ne k\ne i}\frac{n_i^an_j^an_k^a}{\chi_i\chi_j\chi_k}=0 \nonumber\\
n_1^an_2^an_3^a&-&\sum_{i\ne j\ne k\ne i}{m_i^am_j^am_k^a}{\chi_i\chi_j\chi_k}>0 \end{eqnarray}
 where $\chi_i\equiv \left(\frac{{R_2}}{{R_1}}\right)_i$ is the
 complex structure modulus for the $i^{th}$ two-torus.

 At the
intersection of the $U(1)_b$ and $U(1)_c$ D6-branes the chiral matter corresponds to the right handed
neutrinos $N_R$. We insist that there only exist chiral states with such
$U(1)_b$ and $U(1)_c$
charges  that they cannot couple perturbatively in
Yukawa couplings $5\, {\bar 5}N_R$, i.e., $U(1)$ charges
are perturbatively violated for such couplings,
and thus Dirac masses corresponding to them are
not present at this stage. In addition we want to ensure that the D-instanton zero modes induce the Dirac neutrino masses,
while the Majorana  neutrino masses are absent.
   These constraints can be  ensured by the following
% necessary conditions are ensured by
%having, e.g., the following
specific signs for the intersection numbers:
\begin{eqnarray}
I_{ab}&=& 0, \ \ \ \ \ I_{ab'} <0, \nonumber\\
  I_{bc}&=& 0, \ \ \ \ \   I_{bc'} <0, \nonumber\\
   I_{ac}&=&0 , \ \ \ \ \  I_{ac'} >0\, . \label{di}
\end{eqnarray}
resulting in
%With the convention that $I_{ab}< 0$ corresponds to a positive
%number of left-handed chiral
%superfields in $({\bf a}, {\bar b})$, we see
the following
 left-handed  chiral superfield representations:
 $(5_a, 1_b)$,
 %$(1_b,-1_c)$,
  $(1_b,1_c)$, and  $({\bar 5}_a, -1_c)$.
Therefore,
%neither
 $N_R=(1_b,1_c)$
 % nor $(1_b, -1_c)$
  (or any singlets in
 $(b,b')$ and $(c,c')$ sectors with respective charges $\pm 2_b$ and $\pm
2_c$) cannot couple
perturbatively via Yukawa couplings to
$5_a$  and  ${\bar 5_b}$.
% To simplify  the spectrum
%somewhat more
%we require that $I_{bc}=0$, as
%well, and thus focus only on non-perturbative couplings  of  $(5_a,
%1_b)$, $({\bar 5}_a, -1_c)$ and $N_R=
%(1_b,1_c)$ chiral
% left-handed representations.
 We also assume that the states
in
the N=2 sector of $(a,b)$, and $(b,c)$ system are massive by wrapping the
respective D-branes on parallel one-cycles that do not coincide.
 The wrapping numbers
presented in the Table I  comply with the above  intersection number conditions and have the following specific values:
\begin{eqnarray}
I_{ab}&=&  0, \ \ \ \ \  I_{ab'}= -16\nu_2, \nonumber\\
  I_{bc} &=& 0, \ \ \ \ \  I_{bc'}=-16, \nonumber\\
    I_{ac}&=&0 , \ \ \ \ \
I_{ac'} =16\frac{\nu_2}{n_2},
\end{eqnarray}
as are also listed in terms of the multiplicity of the states in the spectrum in Table \ref{sp}.

To generate the desired Yukawa couplings at the non-perturbative level the instanton has to
have a spectrum of zero modes  ensured by the intersection numbers:
\begin{equation}
 I_{aE}=0, \ \  I_{bE}=2, \ \ I_{cE}^{N=2}=1\, . \label{ei}
 \end{equation}
Note that for the non-chiral intersection  with  $I_{cE}=0$, however,  the
N=2  $(c,E)$ sector has one vector pair
of massless modes. To ensure that the  N=2 $(a,E)$ sector does not have massless modes,
the parallel one-cycles on the third two-torus for the $D6_a$-brane  and
the E2-instanton do not coincide. We
therefore ensure the correct structure of the zero modes, i.e., two zero modes
${\overline{\lambda}}_b$ in the representation  $(-1_b,1_E)$ and one vector pair ${\lambda}_c+{\overline
{\lambda}}_c$, $(-1_c,1_E)+(1_c,-1_E)$.

Importantly, D-instanton zero modes (\ref{ei}) {\it cannot generate Majorana masses} for $N_R=(1_b,1_c)$. We have also checked that in the concrete set-up there is no other ${\tilde E}$-instanton that could induce Majorana masses. Such an instanton would have to wrap a rigid three-cycle with the intersection numbers:
 \begin{equation}
 I_{{\tilde E}{\tilde E'}}=0, \ \ I_{E{\tilde E}}=0, \ \
 I_{a{\tilde E}}=0,\ \   I_{b{\tilde E}}=2, \ \ I_{c{\tilde E}}=2\, . \label{etildei}
 \end{equation}
The first three constraints  require $(n^{\tilde E}_3,\ m^{\tilde E}_3)=(0,-1)$, however the last two constraints cannot be satisfied for any  wrapping numbers $(n^{\tilde E}_{1,2},\ m^{\tilde E}_{1,2})$, corresponding to a rigid, supersymmetric three-cycle.

Note  that specific  conditions on the intersection numbers between
D-branes  (\ref{di}) and the intersection numbers of the D-instanton with
D-branes  (\ref{ei}) apply to a general  Type IIA set-up and ensure a
mechanism that generates perturbatively absent   Dirac neutrino masses due
to E2-instantons, which cannot generate  Majorana masses for $N_R$'s.

\begin{figure}[h]
\vskip 1cm
\begin{center}
 \includegraphics[width=0.3\textwidth]{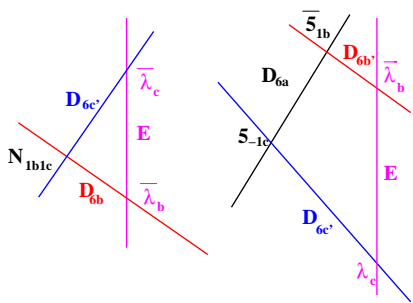}
\end{center}
\caption{\small String disc diagrams  that ensure the absorption of the  fermionic zero modes
$2\times {\bar\lambda}_b$ and ${\lambda_c}+{\overline{\lambda}_c}$.}\label{disk}
\end{figure}

 The contribution to the superpotential  arises from the string amplitudes as shown in
figure \ref{disk}. These four fermionic
zero modes $\lambda$ can be saturated via the two disk
diagrams, thereby generating superpotential contributions  to the Yukawa couplings  ${\bar 5}\, { 5} N_R$
of the type:
\begin{equation}
Y=\exp (-S_{inst})=x\exp (-\frac{2\pi}{\alpha_{GUT}}
\frac{Vol_{E2}}{Vol_{D6a}})
\end{equation}
where $\frac{Vol_{E2}}{Vol_{D6a}}=(6\nu_a)^{-1}$
for the specific local construction. The numerical factor, $x$, is expected
to be of order 1. A more detailed conformal field theory calculation of the
three-point
and four-point disc amplitudes \cite{Cvetic:2003ch,Cvetic:2007ku} emerging
from the geometric  local set
up  in Figure \ref{disk}  could in principle generate additional
world-sheet instanton supression terms, as were explicitly calculated for
the D-instanton induced Majorana masses in \cite{Cvetic:2007ku}.
In addition, further  summation over $Z_2\times Z_2'$ images of the
E-instanton can quantitatively affect $x$, as again was addressed in
\cite{Cvetic:2007ku}.

 Taking
$\nu_2=1$, $\alpha_{GUT} \sim \{25^{-1}, {30}^{-1}\} $  and a VEV of the Higgs doublet $\sim 100 $
GeV  yields neutrino Dirac masses in the range
\begin{equation}
 m_{Dirac}\sim  \ \{ 2\, \times 10^{-3},  0.4\} {\rm \ eV}
\end{equation}
which is a reasonable  regime for the  allowed range for the neutrino masses. Note, however,  that the
case $\nu_2=n_2=3$  would require too  small a value  $\alpha_{GUT}\sim 10^{-2}$ to bring $m_{Dirac}$ to the
$10^{-3}$ eV regime.

In conclusion, we have presented
specific conditions for a concrete  proposal, explicitly implemented
within a local supersymmetric $SU(5)$ construction with intersecting D6-branes,
where string
D-instantons (E2-instantons) can generate  perturbatively absent Dirac
neutrino masses  while Majorana masses remain absent. The
exponentially suppressed coupling can be engineered (without fine-tuning) to yield  Dirac neutrino
masses  in the observed regime $\gtrsim 10^{-3}$ eV. While the concrete  set-up was in the context of Type IIA string
theory,  realizations in the T-dual picture, namely the Type I framework, may be amenable to
constructions of global models on compact Calabi-Yau spaces where string D-instanton couplings are realized within globally consistent models,  \`a la \cite{Cvetic:2007qj}.

\emph{Acknowledgements}
We would like to thank  Timo Weigand  for discussions. We
are especially grateful to  Robert Richter for discussions and
comments  on the manuscript. This research was supported in part by the
National Science Foundation under Grant  PHY-0503584 (P.L.),
the Department of Energy Grant DOE-EY-76-02-3071 (M.C.), the Fay R. and
Eugene L. Langberg Endowed Chair  (M.C.) and by the  Friends of the IAS
(P.L.).
%\newpage

%\newpage

%%%%%%%%%%%%%%%%%%%%%%%%%%%%%%%%%%%%%%%%%%%%%%%%%%%%%%%%%%%%%%%%%
%%%
%%%                     BIBLIOGRAPHY
%%%
%%%%%%%%%%%%%%%%%%%%%%%%%%%%%%%%%%%%%%%%%%%%%%%%%%%%%%%%%%%%%%%%%

%\newpage
%\vskip .75 in

\end{document}